\documentclass{PoS}
\usepackage{cite}
\usepackage{physics}
\usepackage{url}
\usepackage[frozencache,cachedir=minted-files]{minted}
\usepackage{subcaption}

\usepackage{color}
\usepackage{tikz}
\usetikzlibrary{decorations.pathmorphing,decorations.markings}
\usetikzlibrary{decorations.pathreplacing}
\usetikzlibrary{positioning, calc, shapes}
\usetikzlibrary{shapes.geometric, arrows}
    \tikzstyle{startstop} = [rectangle, rounded corners, minimum width=2cm,text centered, draw=black, fill=red!30]
    \tikzstyle{check} = [rectangle, minimum width=2.0cm, text centered, draw=black, fill=blue!30]
    \tikzstyle{iout} = [trapezium, trapezium left angle=-70, trapezium right angle=-110, minimum width=3cm, text centered, draw=black, fill=blue!30]
    \tikzstyle{io} = [trapezium, trapezium left angle=70, trapezium right angle=110, minimum width=3cm, minimum height=1cm, text centered, draw=black, fill=blue!30]
    \tikzstyle{nnpdf} = [rectangle, minimum width=3cm, minimum height=1cm, text centered, draw=black, fill=orange!30]
    \tikzstyle{n3py} = [rectangle, minimum width=3cm, minimum height=1cm, text centered, draw=black, fill=green!30]
    \tikzstyle{n3cpp} = [rectangle, minimum width=3cm, minimum height=1cm, text centered, draw=black, fill=blue!30]
    \tikzstyle{procblue} = [rectangle, minimum width=3cm, minimum height=1cm, text centered, draw=black, fill=blue!30]
    \tikzstyle{fitted} = [rectangle, minimum width=5cm, minimum height=1cm, text centered, draw=black, fill=red!30]
    \tikzstyle{fixed} = [rectangle, minimum width=5cm, minimum height=1cm, text centered, draw=black, fill=blue!30]
    \tikzstyle{arrow} = [thick,->,>=stealth]

\definecolor{darkgreen}{rgb}{0.0, 0.5, 0.13}
\definecolor{darkred}{rgb}{0.55, 0.0, 0.0}
\newcommand{\evolker}{\texttt{evolutionary\_keras}~}
\title{Studying the parton content of the proton with deep learning models}
\ShortTitle{Studying the parton content of the proton with deep learning models}
\author{
    \speaker{Juan M Cruz-Martinez},
    Stefano Carrazza, Roy Stegeman \\
     TIF Lab, Dipartimento di Fisica, Universit\`a degli Studi di Milano and INFN Sezione di Milano \\
        Via Celoria 16, 20133, Milano, Italy
     \\
    E-mail: \email{juan.cruz@mi.infn.it}, \email{stefano.carraza@mi.infn.it}, \email{roy.stegeman@mi.infn.it}
}

\abstract{Parton Distribution Functions (PDFs) model the parton content of the proton. Among the many collaborations which
    focus on PDF determination, NNPDF pioneered the use of Neural Networks to model the
    probability of finding partons (quarks and gluons) inside the proton with a given energy and momentum.  In this
    proceedings we make use of state of the art techniques to modernize the NNPDF methodology and study different models
    and optimizers in order to improve the quality of the PDF: improving both the quality and efficiency of the fits.
    We also present the \evolker library, a Keras implementation of
    the Evolutionary Algorithms used by NNPDF.
}

\FullConference{Artificial Intelligence for Science, Industry and Society, AISIS2019\\	
	October 21-25, 2019\\	
	Universidad Nacional Autonoma de Mexico, Mexico City, Mexico}

\begin{document}

\section{Introduction}
The determination of parton distribution functions (PDFs) is a particular topic which strongly relies on three dynamic
and time dependent factors: new experimental data, higher order theoretical predictions and fitting methodology.
There are two main tasks for PDF fitters such as the NNPDF
collaboration~\cite{AbdulKhalek:2019ihb, Ball:2017nwa}, MMHT~\cite{Harland-Lang:2014zoa} or
CTEQ~\cite{Hou:2019jgw}.
The first of which consists in maintaining and organizing a workflow which incrementally
implements new features proposed by the respective experimental and theoretical communities.  The second task of a PDF
fitter is the investigation of new numerical and efficient approaches to PDF fitting methodology.  While the former
relies almost exclusively on external groups and communities, the latter is under full control of the PDF fitting
collaborations, and in most cases it reflects the differences between them.

In their most generic form, PDFs are functions which depend on an energy scale (usually labelled $Q$) and an energy
fraction ($x$) which represents the fraction of the energy of the proton carried by the parton.
Detailed reviews on the topic can be found in Refs.~\cite{Forte:2013wc,Rojo:2019uip,Ethier:2020way}.

Historically PDFs have been fitted via fixed functional forms which were made more complex as the amount of available data
increased~\cite{Duke:1983gd, Eichten:1984eu, Martin:1987vw, Diemoz:1987xu}.
The work presented in this proceedings is based on the framework developed by the NNPDF collaboration
which introduces for the first time in the PDF determination context the employment of unbiased functional forms
in the form of neural networks to model the parton content of the proton~\cite{Forte:2002fg,Ball:2008by}.

\section{Neural Networks PDF: NNPDF}
The target goal in PDF determination is the optimization of the $\chi^{2}$ of the fit.
This is a problem well suited for machine learning techniques, as the functional form is unknown while the input and
output are well defined.
The details for the current implementation of the NNPDF methodology can be found in~\cite{Ball:2017nwa, Ball:2014uwa}.
An exhaustive overview is beyond the scope of this proceedings,
in what follows we outline some of the most relevant aspects of the NNPDF methodology used in the last main releases
(3.0 and 3.1)
as a baseline for our current discussion.

In the NNPDF methodology fits are parameterized at a reference scale $Q_0$ and expressed in terms of a neural network
whose outputs correspond to a set of basis functions.
The neural network consists of a fixed-size feedforward multi-layer perceptron.
The input node ($x$) is split by the first layer on the pair $(x,\log(x))$.
The hidden layers use the sigmoid activation function while the output node is linear.
The functional form of this model is:
\begin{equation}
    f_i(x,Q_0) = A_i x^{-\alpha_i} (1-x)^{\beta_i} {\rm NN}_i(x), \label{eq:PDFdefinition}
\end{equation}
where ${\rm NN}_{i}$ is the output of the neural network corresponding to a given flavour $i$,
usually expressed in terms of the PDF evolution basis $\{g,\,\Sigma,\,V,\,V_3,\,V_8,\,T_3,\,T_8,\,c^+\}$.
$A_i$ is an overall normalization constant which enforces sum rules and $x^{-\alpha_i} (1-x)^{\beta_i}$ is
a preprocessing factor which controls the PDF behaviour at small and large $x$.
Both the $\alpha_i$ and $\beta_i$ parameters are randomly selected within a defined range for each replica at the
beginning of the fit and kept constant thereafter.

Unlike usual regression problems, where during the optimization procedure the model is compared directly to the training
input data, in PDF fits the theoretical predictions are constructed through the convolution operation per data point
between a fastkernel table (FK)~\cite{Ball:2010de,Bertone:2016lga} and the PDF model.
For DIS-like processes the convolution is
performed once, while for hadron collision-like processes PDFs are convoluted twice (once per hadron).

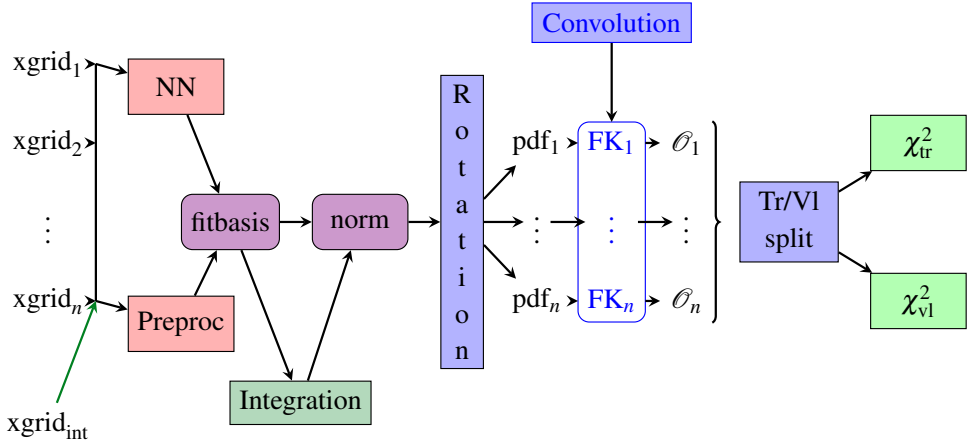
\begin{figure}[tb]
    \centering
    \resizebox{0.85\textwidth}{!}{%
    \begin{tikzpicture}[node distance = 1.0cm] \small
        \node (x1) {$\text{xgrid}_{1}$};
        \node[below of = x1] (x2) {$\text{xgrid}_{2}$};
        \node[below of = x2] (xd) {\vdots};
        \node[below of = xd] (xn) {$\text{xgrid}_{n}$};
        \node[below = 1cm of xn] (xint) {$\text{xgrid}_{\text{int}}$};

        \node[fitted, right = 1.0cm of x1.south, minimum width=1.2cm, minimum height=0.7cm]
                (pdf) {NN};
        \node[fitted, right = 1.0cm of xn.south, minimum width=1.2cm, minimum height=0.7cm]
                (preproc) {Preproc};

        \node[startstop, fill=violet!40, minimum width=1.2cm, minimum height=0.7cm, right = 1.5cm of xd.east]
                (fitbasis) {fitbasis};
        \node[startstop, fill=violet!40, minimum width=1.2cm, minimum height=0.7cm, right = 0.4cm of fitbasis]
                (normalizer) {norm};

        \node[fixed, minimum width=0.3cm, right = 0.4 of normalizer, text width=0.3cm]
                (rotation) {R o t a t i o n};

        \node[right = 0.5cm of rotation] (pd) {\vdots};
        \node[above of = pd] (p1){pdf$_{1}$};
        \node[below of = pd] (pn) {pdf$_{n}$};

        \node[blue,right = 0.6cm of pd] (fd) {\vdots};
        \node[blue,above of = fd] (f1) {FK$_{1}$};
        \node[blue,below of = fd] (fn) {FK$_{n}$};
        \node[procblue, minimum width = 2cm, minimum height=0.5cm, draw=blue, above = 1cm of f1] (convolution) {\color{blue}Convolution};
        \draw[draw=blue, rounded corners] (f1.north west) rectangle (fn.south east);

        \node[right = 0.6cm of fd] (od) {\vdots};
        \node[above of = od] (o1) {$\mathcal{O}_{1}$};
        \node[below of = od] (on) {$\mathcal{O}_{n}$};

        \node[fixed, right = 0.5cm of od, minimum width = 1.2cm, text width=1cm, minimum height=0.7cm]
                (trvl) {Tr/Vl split};
        \coordinate[right = 1cm of trvl] (cp);
        \node[n3py, above of = cp, minimum width = 1.2cm, minimum height=0.7cm]
                (chi2tr) {$\chi^{2}_\text{tr}$};
        \node[n3py, below of = cp, minimum width = 1.2cm, minimum height=0.7cm]
                (chi2vl) {$\chi^{2}_\text{vl}$};

            \node[procblue, fill=darkgreen!30, minimum width=1.0cm, minimum height=0.5cm] at ($(preproc) + (1.5, -1.0)$)
                (integration) {Integration};

        \draw[arrow] (pdf) -- (fitbasis);
        \draw[arrow] (preproc) -- (fitbasis);
        \draw[arrow] (fitbasis) -- (normalizer);
        \draw[arrow] (normalizer) -- (rotation);
        \draw[arrow] (trvl) -- (chi2tr);
        \draw[arrow] (trvl) -- (chi2vl);
        \draw[arrow] (integration) -- (normalizer);
        \draw[arrow] (fitbasis) -- (integration);
        \draw[arrow] (rotation) -- (p1);
        \draw[arrow] (rotation) -- (pd);
        \draw[arrow] (rotation) -- (pn);
        \draw[arrow] (p1) -- (f1);
        \draw[arrow] (pd) -- ($(fd) + (-0.33, 0.0)$);
        \draw[arrow] (pn) -- (fn);
        \draw[arrow] (f1) -- (o1);
        \draw[arrow] ($(fd)+(0.33,0.0)$) -- (od);
        \draw[arrow] (fn) -- (on);
        \draw[arrow] (convolution) -- (f1.north);

        \draw[decorate, decoration={brace}, thick] (o1.north east) -- (on.south east);


        \coordinate (a1) at ($(x1) + (0.6, 0.0)$);
        \draw[arrow] (x1) -- (a1);
        \draw[arrow] let
                \p1 = (a1), \p2 = (x2) in
                (x2) -- (\x1, \y2);
        \draw[arrow] let
                \p1 = (a1), \p2 = (xn) in
                (xn) -- (\x1, \y2);

        \draw[thick] let
                \p1 = (a1), \p2 = (xn) in
                (a1) -- (\x1, \y2);

        \draw[arrow] (a1) -- (pdf);
        \draw[arrow] let
                \p1 = (a1), \p2 = (xn) in
                (\x1, \y2) -- (preproc);

        \draw[arrow, darkgreen] let
                \p1 = (a1), \p2 = (xn) in
                (xint) -- (\x1, \y2);
    \end{tikzpicture}
    }
    \caption{Diagrammatic view of the NNPDF fitting model as implemented in the \texttt{n3fit} code. Each different
    block is set as a different Keras layer. The red squared blocks correspond to
blocks with potentially fittable parameters while other blocks represent parameters or transformations which are fixed
during the fit.}
    \label{fig:n3fit}
\end{figure}

The optimization procedure consists in minimizing the loss function
\begin{equation}
    \text{Loss} = \chi^2 = \sum_{i,j}^{N_{\rm dat}} (D-P)_i \sigma_{ij}^{-1} (D-P)_j, \label{eq:chi2}
\end{equation}
where $D_i$ is the $i$-th artificial data point from the training set, $P_i$ is the convolution product between the
fastkernel tables for point $i$ and the PDF model, and $\sigma_{ij}$ is the covariance matrix between data points $i$
and $j$ (following the $t_0$ prescription defined in appendix of Ref.~\cite{Ball:2012wy}).
The covariance matrix can include both experimental and theoretical components as presented in Ref.~\cite{AbdulKhalek:2019bux}.
As a summary, in Fig.~\ref{fig:n3fit} a schematic view of the code is represented.

Up to the current NNPDF release (3.1 at the time of writing of these proceedings, Ref.~\cite{Ball:2017nwa})
only genetic algorithms had been used as the optimizer for the weights and biases of the neural network.
The optimization procedure makes use of the Nodal Genetic Optimizer defined in Ref.~\cite{Ball:2014uwa} and can be summarized as follows:
the weights of the neural network are initialized using a random gaussian distribution and checking that sum rules are satisfied.
From that first network 80 mutant copies are created based on a given mutation probability and update size.
The training procedure is fixed to $3\cdot10^{4}$ iterations
and the best mutant is determined using a look-back algorithm which stores the best weights for the lowest validation loss
value.

The experimental data used in the PDF fit is preprocessed according to a cross-validation strategy based on randomly
splitting the data for each replica into a training set and a validation set.
The optimization is then performed on the training set while the validation loss is monitored to reduce overlearning.

The codebase driving this methodology has up to now been a fully in-house C++ program.
Such an in-house codebase allowed for a more fine grained control of the details of the methodology but, as a
consequence, the
implementation of new features and studies were hindered by the extra amount of effort for both developers and maintainers.

\section{A new methodology: \texttt{n3fit}}
The advances in the field of machine learning have worked their way into the field of PDFs in the form of a new methodology which inherits directly from
NNPDF and has been named \texttt{n3fit}\footnote{We follow the well known trend in High Energy Physics which relates the
quality of a calculation to the multiplier of the letter \texttt{n} in their name.}.
This new methodology has been presented for the first time in~\cite{Carrazza:2019mzf}.
The codebase for \texttt{n3fit} has been written in Python, with Keras~\cite{chollet2015Keras} and
Tensorflow~\cite{tensorflow2015:whitepaper} as the engine behind the training and construction of the neural network.
Other engines can be implemented as most of the codebase is library agnostic.

This new methodology has successfully introduced several changes to the NNPDF methodology and allowed for an
improvement of the research results and workflow~\cite{Carrazza:2019mzf,Carrazza:2019agm}.
Here we focus on the substitution of the genetic algorithm for an optimization based on gradient descent and on a
comparison with previous results.

\subsection{Optimization algorithm: from genetic algorithms to gradient descent}
The default minimizer of the C++ version of the NNPDF fitting code (\texttt{nnfit})
is the aforementioned Nodal Genetic Algorithm (or NGA).
The NGA has been used, for instance, in NNPDF3.0~\cite{Ball:2014uwa} and NNPDF3.1~\cite{Ball:2017nwa}.
Other evolutionary optimizers have also been implemented in \texttt{nnfit}, such as
Covariance Matrix Adaptation Evolution Strategy (CMA) which was used in the release of hadronic fragmentation
functions~\cite{Bertone:2017tyb,Carrazza:2017udv}.

Newer releases by the NNPDF collaboration explore the usage of deterministic minimizers.
The nuclear PDF set nNNPDF1.0~\cite{AbdulKhalek:2019mzd} make use of the ADAM gradient descent optimizer~\cite{Kingma:Adam}
which is closely related to RMSProp~\cite{Tieleman2012} and Adagrad~\cite{Duchi:Adagrad}.
The full NNPDF fitting mechanism has also been upgraded to support gradient descent optimizers
and it has been implemented in \texttt{n3fit}~\cite{Carrazza:2019mzf}.
The \texttt{n3fit} code supports the usage of the Adagrad, RMSProp and ADAM
minimizers as well as any optimizer compatible with the Keras library.

\begin{figure}[h]
    \centering
    \begin{subfigure}[t]{0.48\textwidth}
        \includegraphics[width=\textwidth]{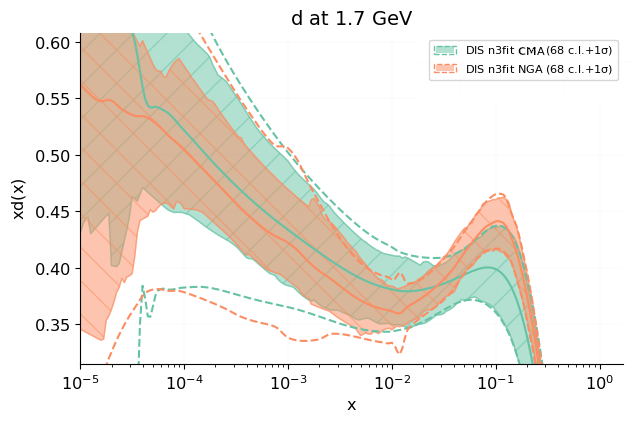}
        \caption{Comparison of the NGA algorithm (orange) and the CMA algorithm (green). The CMA produce a smoother
        curve at the cost of significantly longer running times (up to a factor of two in our tests).}
        \label{fig:cmavsga}
    \end{subfigure}
    \hfill
    \begin{subfigure}[t]{0.48\textwidth}
        \includegraphics[width=\textwidth]{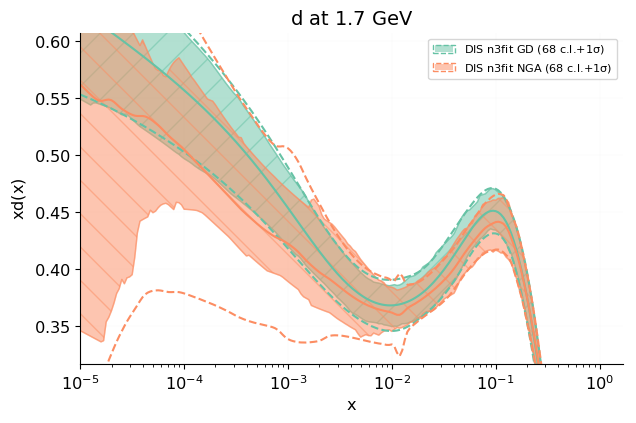}
        \caption{Comparison of the NGA algorithm (orange) and a gradient descent algorithm (green) for the d quark inside the proton. The better stability of the deterministic optimizer produces a smoother error band and central value all across the considered ranges. }
        \label{fig:gavsgd}
    \end{subfigure}
    \caption{Comparison of different algorithms in \texttt{n3fit} in DIS only fit. Note that all of them produce
    compatible result within their error bands in the considered ranges. Plots are generated with \texttt{reportengine}~\cite{zahari_kassabov_2019_2571601}.}
\end{figure}

We do a thorough exploration of algorithms previously used by the NNPDF collaboration and gradient descent based
algorithms.
In Fig.~\ref{fig:cmavsga} we observe that the CMA algorithm produces smoother results than the NGA and thus
convergence is improved: the number of outliers is reduced and postfit selections are less aggressive.
We find, however, longer running times, which makes it unsuitable for our purposes.

For this particular problem we have access to the gradient of the model, as such Gradient Descend (GD) based algorithms
can be expected to outperform evolutionary strategies. 
In Fig.~\ref{fig:gavsgd} we present a comparison between a GD based algorithm and the NGA used in previous releases of
NNPDF.
We find that using SGD results in features similar to those we observe when using CMA, that is, smoother replicas and better convergence.
At the same time, the performance is also improved: computing times are greatly reduced. 
As a consequence, new lines of investigation appear which would not have been possible before, such as hyperparameter
scans~\cite{Bergstra:2013, Carrazza:2019mzf}.

The gradient descent implementation can comfortably be run in a consumer-grade desktop computer
while the NGA instead has been run on a cluster with the aid of
\texttt{pyHepGrid}~\cite{cruz_martinez_juan_2019_3233862}.

The Keras library does not include evolutionary strategies which has motivated us to develop a new library:
\evolker~\cite{juan_cruz_martinez_2020_3630340}.
This library implements evolutionary strategies used by NNPDF retaining full compatibility with any Keras model.
The fit shown in Fig.~\ref{fig:gavsgd} based on the DIS dataset from NNPDF3.1 and labelled GA use the NGA algorithm described
in~\cite{Ball:2014uwa} as implemented in \evolker and it is compared against the Adadelta~\cite{DBLP:adadelta} optimizer
implemented in the standard Keras library.
The engine behind both fits is \texttt{n3fit}.

\subsection{Comparison with old methodology}

\begin{figure}[h]
    \centering
    \includegraphics[width=0.49\textwidth]{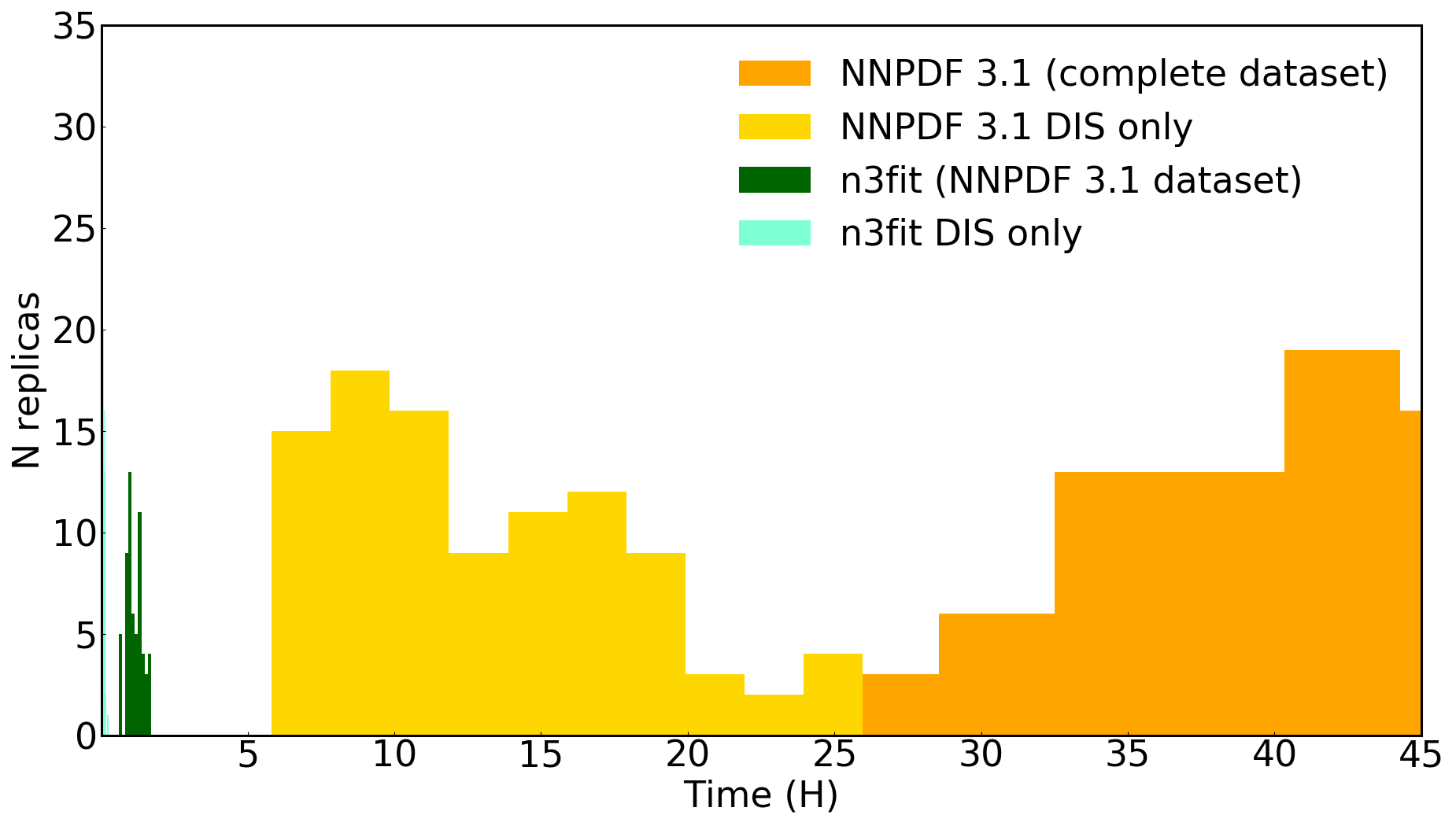}
    \includegraphics[width=0.49\textwidth]{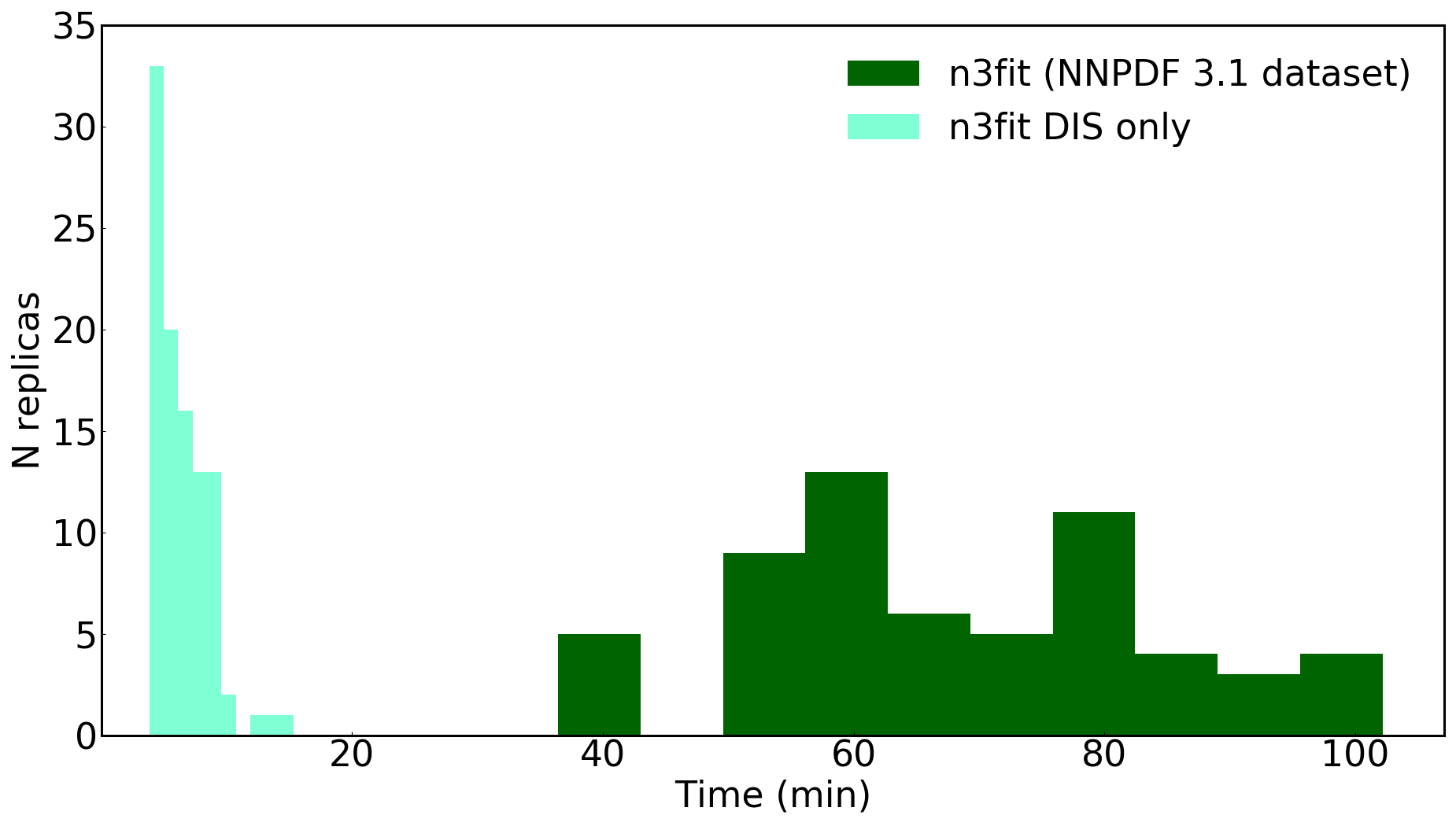}
    \caption{Distribution of computing times for a full DIS and global fit for the NNPDF 3.1 methodology (yellow) and \texttt{n3fit} (green). While for the old methodology the computing time for one replica could take days, the new code achieves similar fit goodness in a scale of hours.}
    \label{fig:times}
\end{figure}

From a technical perspective, one of the most relevant advantages introduces by \texttt{n3fit} is the reduction
of the computational time required to obtain PDF sets based on the NNPDF3.1 dataset~\cite{Ball:2017nwa}.
These improvements are in part due to a combination of the new minimizer based on gradient descent and the usage of multi-threading CPU
calculations when executing the TensorFlow graph model.
In Fig.~\ref{fig:times} we show the distribution of computing time in a per-replica basis
(each fit is composed of up to 100 replicas).
Note that while a global NNPDF fit with the 3.1 methodology could take up to two days per replica, the
same NNPDF fit with the new methodology only takes a few hours.
Further attempts to reduce these figures are currently underway with the implementation of custom TensorFlow operators
and the usage of Graphical Processing Units (GPUs) for the FK convolution~\cite{Carrazza:2019agm}.

\begin{figure}[h]
    \centering
    \includegraphics[width=0.49\textwidth]{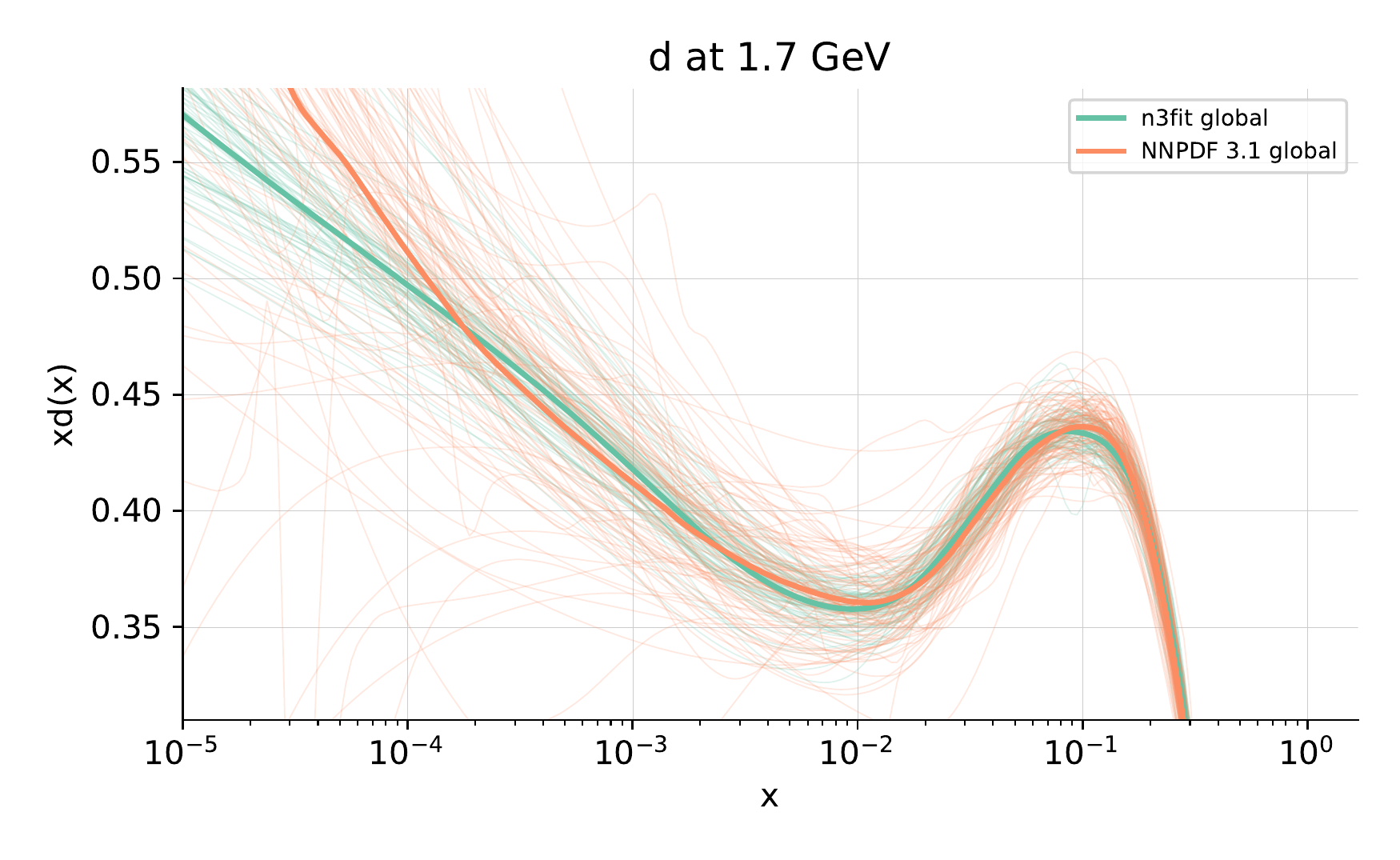}
    \includegraphics[width=0.49\textwidth]{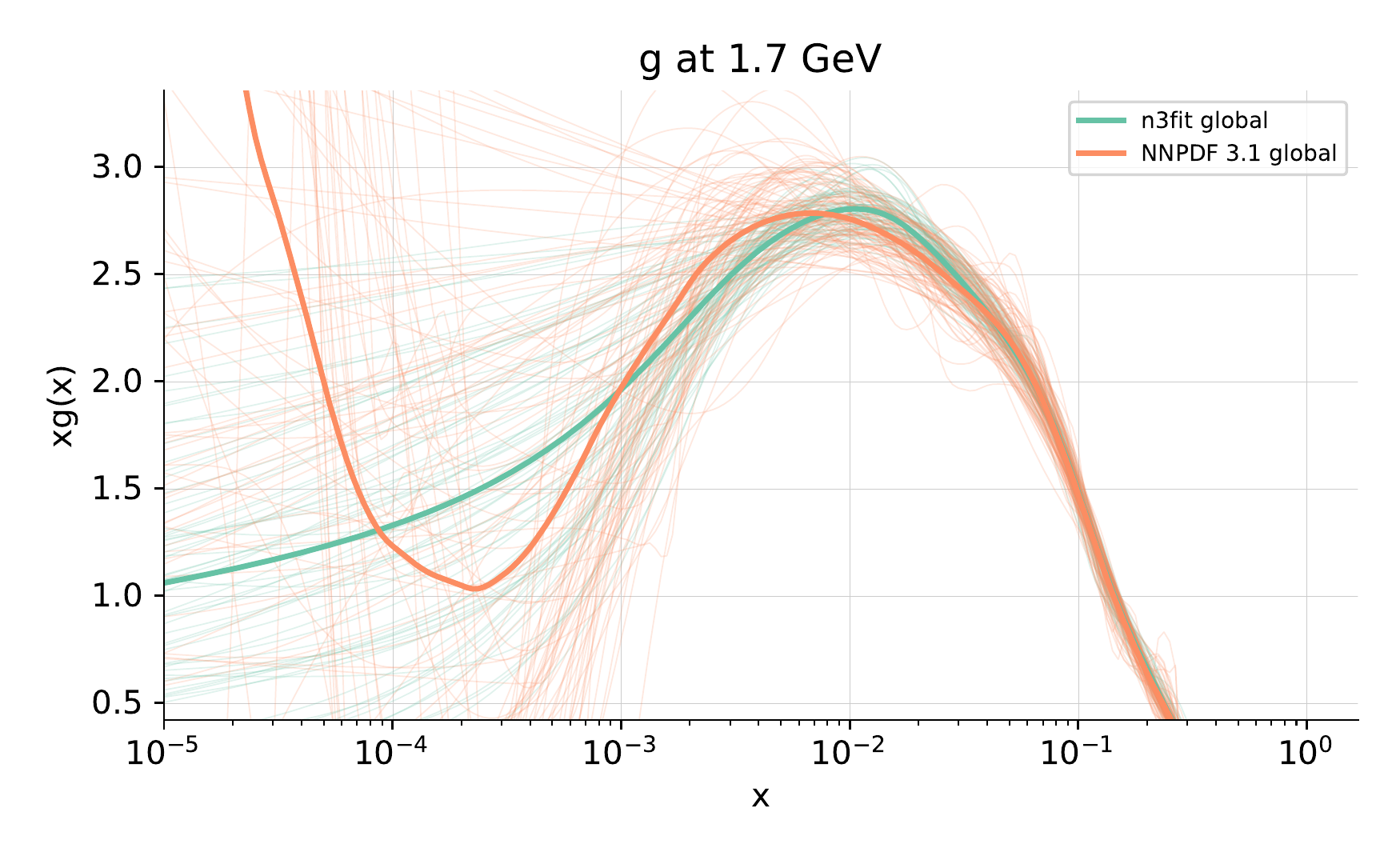}
    \caption{Comparison between the results obtained using the old methodology (orange) and the new \texttt{n3fit}
    methodology (green) for two of the partons found inside the proton, the d quark (left) and the gluon (right). Note
that the results are compatible across the entirety of the studied range with overlapping central values in the region where most of the data is concentrated.}
    \label{fig:fitcompare}
\end{figure}

The number of PDF replicas that are necessary for a full fit is also reduced due to the better replica-by-replica
stability that gradient descent
methods produce for this problem with respect to genetic algorithms.
This stability is also apparent when
we plot the PDF replicas themselves as seen in Fig.~\ref{fig:fitcompare} where in green we show the results of
the new methodology while in orange those of the methodology used up to NNPDF 3.1.
Nevertheless, the central value only shows a different behaviour in the extrapolation error
(where the lack of data and the size of the errorbar make the difference irrelevant).
In the data region \texttt{n3fit} shows much smoother lines.

\section{A modern Genetic Algorithm library: \evolker}
As part of the current research a new library has been developed which aims to implement the evolutionary strategies
used by the NNPDF collaboration.
The library has been named \\
\evolker\cite{juan_cruz_martinez_2020_3630340} and
the code is publicly available in the N3PDF repository at \\
\href{https://github.com/N3PDF/evolutionary_keras}{https://github.com/N3PDF/evolutionary\_keras}.
It can easily be installed with the \texttt{pip} package manager.
\begin{minted}{bash}
    ~$ pip install evolutionary-keras 
\end{minted}

It is also included in the \texttt{conda-forge} repository so it can be installed for Anaconda projects.
\begin{minted}{bash}
    ~$ conda install -c conda-forge evolutionary_keras
\end{minted}

Documentation for the library is found at \href{https://evolutionary-keras.readthedocs.io}{https://evolutionary-keras.readthedocs.io} where all available optimizer are
listed and some examples of usage can be found.

The library consists of a module which extends the Keras Model class and implements optimizers based on
Genetic Algorithms, currently includes the aforementioned NGA and CMA evolutionary strategies.
Compatibility with Keras is retained and it can be used as any other Keras optimizer.

In order to use the capabilities of \evolker in a project, it is necessary to use the model-classes provided. These classes
inherit from the Keras model class and transparently defer to them whenever a Gradient Descent algorithm is used.

Below we present one example in which a Keras model is created with some input and output
tensors previously defined.
Those familiar with Keras will immediately see the similarities.

\begin{minted}[linenos]{python}
    from evolutionary_keras.models import EvolModel
    my_model = EvolModel(input_layer, output_layer)
    from evolutionary_keras.optimizers import NGA
    my_nga = NGA(population_size = 42, mutation_rate = 0.2)
    my_model.compile(optimizer = my_nga, loss = 'mean_squared_error')
    my_model.fit(x = input_data, y = output_data, epochs = 10)
\end{minted}

In summary, any Keras \texttt{Model} must become an \texttt{EvolModel}, but it can be instantiated with the same syntax
as can be seen in line 2.
From that point onwards the usage is unchanged with respect to a normal Keras project.
In line 4 the genetic optimizer NGA is instantiated with some initial parameters and then compiled into the model
together with one of the default loss functions included in Keras.
For an up to date list of included algorithms, please consult the documentation.

\acknowledgments

The authors are supported by the European Research Council under the European Unions Horizon 2020 research and
innovation Programme (grant agreement number 740006). S.C. and J.CM are also supported by the UNIMI Linea 2A grant ``New
hardware for HEP''.

\bibliographystyle{../JHEP}
\bibliography{../blbl}

\end{document}